\documentclass[a4paper,12pt]{article}
\usepackage{amsmath}
\usepackage[dvips]{graphicx}

\begin{document}
\author{Andrzej Ostruszka$^1$ and Karol \.Zyczkowski$^2$\\[2mm]
\parbox{.9\textwidth}{\small%
$^1$Instytut Fizyki im. Mariana Smoluchowskiego,
    Uniwersytet Jagiello\'nski,\\
    ul. Reymonta 4,  30-059 Krak\'ow, Poland\\[2mm]
$^2$Centrum Fizyki Teoretycznej PAN, \\
    al. Lotnik\'ow 32/44, 02--668 Warszawa, Poland \\[2mm]
e-mail addresses: \texttt{ostruszk@if.uj.edu.pl},
         \texttt{karol@cft.edu.pl}
}}
\title{Spectrum of the Frobenius--Perron\\
operator for systems with\\
stochastic perturbation}
\date{May 17, 2001}
\maketitle
\section{Introduction}
Consider a discrete dynamical system $f$ acting on a phase space $\Omega$.
Time evolution of a classical density $\rho$ (a probability distribution
function on~$\Omega$) is governed by the Frobenius--Perron operator $M$
\cite{Schuster, Gaspbook}
\begin{equation}
\rho'(y)=(M\rho)(y)=\int_\Omega\delta(f(x)-y)\rho(x)dx,\quad
    \mathrm{where}\ \int_\Omega\rho(x)dx = 1\,.
\label{eq:FP}
\end{equation}
Its spectral properties are of considerable interest, since they influence
properties of the correlation function \cite{Driebe}, which may be measured in
a physical system. Spectrum of the FP-operator was calculated for classical
dynamical systems including tent map, Bernoulli shift, baker map
\cite{Hasegawa,Gaspard,Fox} and the so--called ``four legs'' map
\cite{Dellnitz}. However, results obtained may depend on the choice of the
function space FP-operator operates in. Such a case was explicitly
demonstrated by Antoniou \emph{et al.} \cite{Anton}, in which two different
spectra for a certain random dynamical system were found. One might ask
therefore, which of those constructions lead to physically meaningful
results.

In all physical situations the system under consideration is inevitably
subjected to a stochastic perturbation which we usually want to reduce as much
as we can. The idea we want to present here is that the presence of a
perturbation of a small amplitude may choose one particular space of functions
and thus define a ``physical'' decomposition of the FP-operator, without
ambiguities related to the choice of the space of the eigenstates. Note that
the natural invariant measure (so--called SRB measure \cite{Katok}) of a
classical dynamical system may be defined in a slightly analogous way, as the
unique invariant measure stable with respect to stochastic perturbation.
Moreover, it is instructive to recall the quantum mechanical problem of
calculating corrections to degenerated energy levels via perturbation theory.
A priori none of the basis in subspace of degenerated states is distinguished.
However, one may distinguish a certain basis by applying an arbitrary
perturbation and later decreasing its amplitude to zero.

\section{Model system}

We analyze a model system introduced in \cite{OPSZ}
which allows one for an \emph{exact} representation of the FP-operator
describing dynamical system with noise as a finite dimensional matrix. The
construction proceeds as follows.

For simplicity we consider a discrete dynamical system $f$ acting on the
interval $\Omega = [0,1)$ and subjected to an additive noise (with periodic
boundary conditions)
\begin{equation}
x_{n+1}=f(x_n)+ \xi_n\quad\pmod1\,,
\label{eq:langev}
\end{equation}
where $\xi_1, \xi_2, \dots$ are independent random variables fulfilling
\begin{eqnarray*}
\mathrm{stationarity}&&\quad{\cal P}(\xi_n) = {\cal P}(\xi)\,,\\
\mathrm{zero\ mean}&&\quad\langle \xi_n \rangle = 0\,,\\
\mathrm{finite\ variance}&&\quad\langle \xi_n \xi_m \rangle = 
    \sigma^2 \delta_{mn}\,.
\end{eqnarray*}
We choose for a noise for which the probability of transition from $x$ to $y$
(${\cal P}(x,y) = {\cal P}(\xi)$) is
\begin{subequations}
\begin{eqnarray}
\mathrm{homogeneous}& &\quad{\cal P}(x,y) \equiv {\cal P}(x-y), \\
\mathrm{periodic}& &\quad{\cal P}(x,y) \equiv {\cal P}(x+1,y)
    \equiv {\cal P}(x,y+1), \\
\mathrm{decomposable}& &\quad{\cal P}(x,y) =
    \sum_{l,r=0}^{N}A_{lr}u_r(x)v_l(y), \label{eq:expand}
\end{eqnarray}
\end{subequations}
where $A = (A_{lr})_{l,r = 0, \dots ,N}$ is a real matrix of finite size
($N+1$) of expansion coefficients and $(u_r)_{r=0,\dots,N},\
(v_l)_{l=0,\dots,N}$ are two sets of linearly independent real valued
functions.

All these conditions are satisfied by the trigonometric noise
\begin{equation}
\label{eq:trignoise}
{\cal P}_N(x,y) = {\rm C_N}\cos^N(\pi[x-y])\,,
\end{equation}
where $N$ is even and the normalization constant
${\mathrm C_N}=\frac{\sqrt\pi\Gamma(N/2+1)}{\Gamma(N/2+1/2)}$ assures
\mbox{$\int_0^1{\cal P}_N(x)dx=1$}. The noise
strength may be parametrized either by $N$ or by its variance \cite{OPSZ}
\begin{equation}
\sigma_N^2 = \frac{1}{2\pi^2}\Psi'(N/2+1)
  = \frac{1}{2\pi ^{2}}\left(
    \sum_{k=N/2+1}^{\infty }\frac{1}{k^{2}}\right)
  = \frac{1}{12}-\frac{1}{2\pi^2}\sum_{k=1}^{N/2}\frac{1}{k^2},
\end{equation}
where $\Psi'$ stands for the derivative of the digamma function \cite{GR65}.
Asymptotically
$\sigma_N^2\sim1/N\mathop{\rightarrow}\limits_{(N\rightarrow\infty)}0$.
In this case $A_{lr} = \binom{N}{l}\delta_{lr}$, while
$u_l(x)=v_l(x)=\sin^l(\pi x)\cos^{N-l}(\pi x)$.
\begin{figure}
\centering
\includegraphics[width=0.6\textwidth]{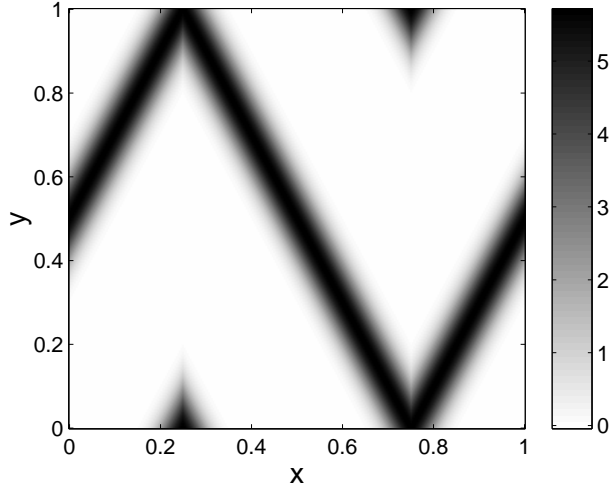}
\caption{Transition kernel ${\cal P}_N(f(x),y)$ for the map defined by
(\ref{eq:antikel_map}) with $\epsilon=0$ and the noise strength characterized
by $N=20$, which gives $\sigma\approx0.0694$.} \label{fig:noise}
\end{figure}

For a deterministic system, the action of the FP-operator is described by
(\ref{eq:FP}). In the presence of a stochastic perturbation this
equation needs to be modified,
\begin{equation}
\label{eq:FPnoise}
\rho'(y) = \int{\cal P}(f(x),y)\rho(x)dx\,.
\end{equation}
Due to the decomposition property (\ref{eq:expand}) we may write
\begin{eqnarray}
\rho'(y)
 & = & \sum_{l,r=0}^N A_{lr} \int_0^1 u_r(f(x)) v_l(y) \rho(x) dx \nonumber\\
 & = & \sum_{r=0}^N \left[\int_0^1 u_r(f(x))\rho(x)dx\right] \tilde{v}_{r}(y)\,,
\end{eqnarray}
where $\tilde{v}_{r} = \sum_{l=0}^N A_{lr}v_l$. Thus any initial density is
projected by the FP-operator into the vector space spanned by the functions
$\tilde{v}_{r}$. Eventually, we can represent the action of this operator by
a matrix of size $N+1$ acting on the vector of expansion coefficients. Writing
$\rho(x) = \sum_l \alpha_l\tilde{v}_l(x)$ we obtain
\begin{equation}
\rho'(y) = \sum_k\tilde{v}_k(y)\alpha_k' =
    \sum_{kl}\tilde{v}_k(y)\,\mathbf{D}_{kl}\,\alpha_l\,,
\label{eq:evolution}
\end{equation}
where $\mathbf{D}_{kl}=\int dx\,u_k(f(x))\,\tilde{v}_l(x)$.
It is worth emphasizing that the matrix representation of the FP-operator
for the system with noise is \emph{finite dimensional}, while the
FP-operator connected with the deterministic case acts in an infinite
dimensional space.

\section{Examples}

In \cite{Gasp} it was conjectured that for continuous dynamics the spectrum of
the system with noise should tend to the spectrum of the operator describing
the deterministic system, if the amplitude of noise tends to zero. This
correspondence is based on solving eigenvalue problem for the Foker--Planck
equation -- which a priori is not a simpler task than calculating the spectrum
of the Frobenius--Perron operator. In our approach we can find approximation
of the exact spectrum just by matrix diagonalization. The larger dimension of
the matrix $\mathbf{D}$, the smaller variance $\sigma^2$ of the noise, and the
better the approximation of the spectrum of the deterministic system. In order
to show, how the spectrum of the Frobenius--Perron operator could be
approximated by our procedure we calculate it for a simple one--dimensional
map.

First of all we have to consider stability of the spectrum if the
system is subjected to a stochastic perturbation. To address this problem it
is useful to introduce a notion of \emph{essential spectrum}
\cite{Keller,Keller01}. It is a part of the spectrum contained in the
smallest disc of radius $r$ such that all eigenvalues outside it are isolated
and of finite multiplicity\footnote{For linear piecewise expanding 1D maps it
was shown in \cite{Keller84} that
$r=\lim_{n\rightarrow\infty}\sqrt[n]{1/\lambda_{f^n}}$ where $\lambda_f$ is
the smallest absolute value of the derivative of $f$.}.
\begin{figure}
\centering
\includegraphics[width=0.6\textwidth]{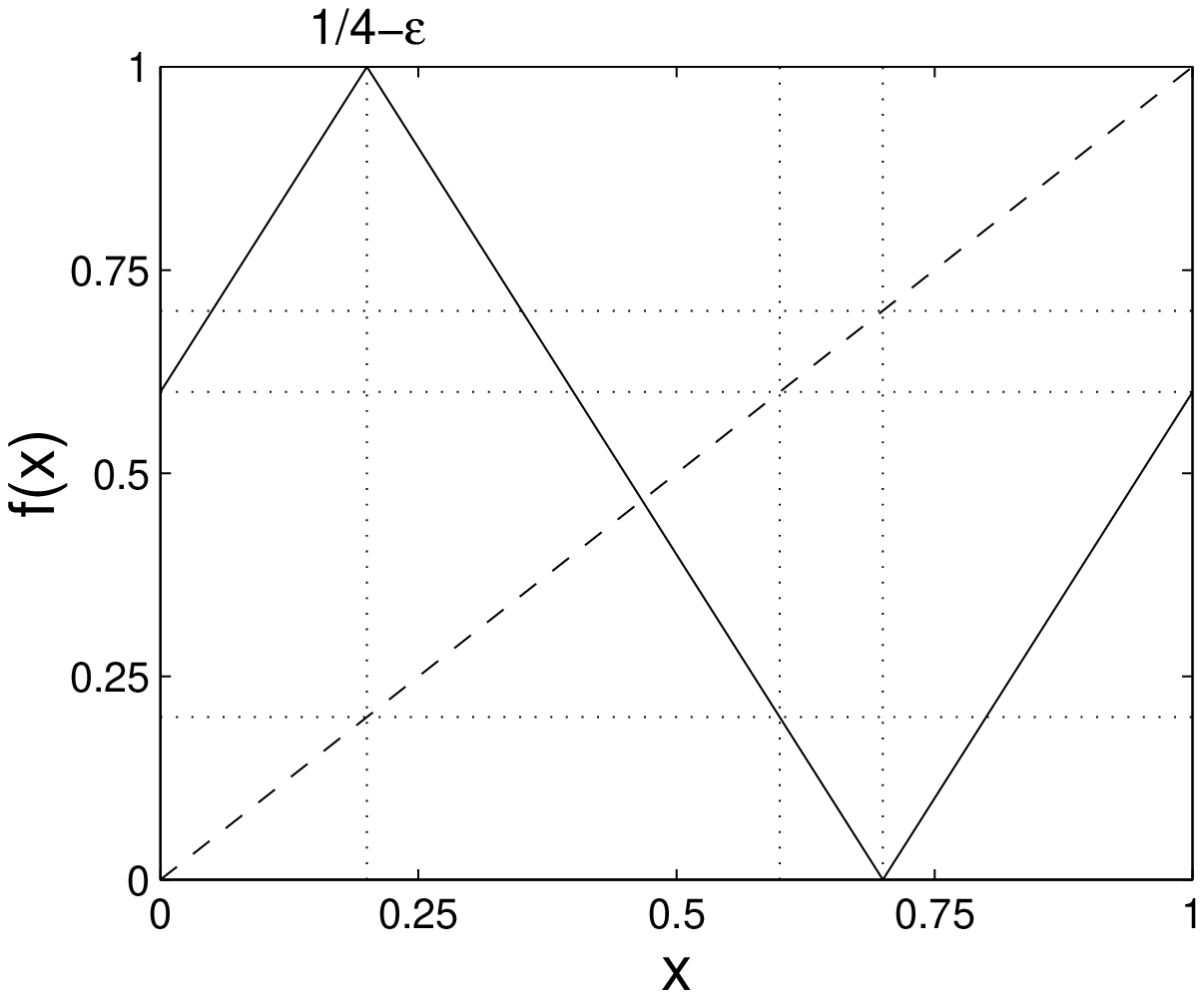}
\caption{Map given by Eq.~(\ref{eq:antikel_map}) for $\epsilon=1/20$}
together with its Markov partition denoted by dotted lines.
\label{fig:antikel}
\end{figure}
Blank and Keller have shown \cite{Keller} that for piecewise expanding maps
(without periodic points, for which the derivative $f'$ is not defined)
subjected to any local perturbation, eigenvalues outside essential spectrum
are close to those of the deterministic system.

We analyze a family of piecewise expanding maps which have isolated
eigenvalues outside the essential spectrum
\begin{equation}
\label{eq:antikel_map}
f(x)=\left\{
\begin{array}{ll}
2(x+\epsilon)+1/2 & x<1/4-\epsilon \\
3/2-2(x+\epsilon) & x\in[1/4-\epsilon, 3/4-\epsilon) \\
2(x+\epsilon)-3/2 & x\geq 3/4-\epsilon
\end{array}\right.\ .
\end{equation}
and subject it to the perturbation (\ref{eq:trignoise}) (non local in sense of
\cite{Keller}).
Since the absolute value of the slope $|f'(x)|$ is constant and equal
to~2, the radius of the essential spectrum equals $1/2$ independently of the
parameter $\epsilon$. For $\epsilon=0$ there are two eigenvalues with modulus
$1$ ($\pm1$) -- the system separates into two subsystems
$\Omega_1=(0,\frac12)$ and $\Omega_2=(\frac12,1)$ which exchange positions
with each other under every iteration of the map. This fact is related to the
presence of $\lambda_2=-1$ in the spectrum. If we increase $\epsilon$ then the
modulus of the negative eigenvalue will decrease. However, for sufficiently
small $\epsilon$ the subleading (with second largest modulus) eigenvalue is
still located outside the essential spectrum so we expect to approximate it
with our procedure. For certain values of $\epsilon$ we may construct Markov
partitions of $\Omega$, write down the corresponding stochastic transition
matrix $\mathbf{T}$ and find analytically its subleading eigenvalue
$\lambda_2$ which determines the rate of convergence to the equilibrium.
Fig.~\ref{fig:antikel} presents map~(\ref{eq:antikel_map}) for
$\epsilon=1/20$ together with its Markov partition. 
\begin{figure}
\centering
\includegraphics[width=0.6\textwidth]{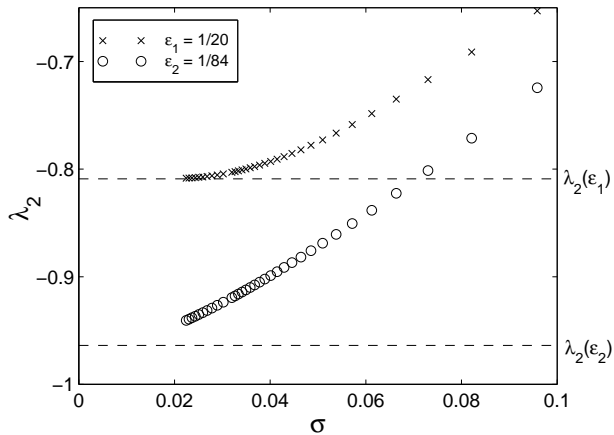}
\caption{Subleading eigenvalue as a function of noise width $\sigma$ for map
defined by (\ref{eq:antikel_map}) plotted for $\epsilon=1/20$ and
$\epsilon=1/84$ together with the deterministic limit represented by
horizontal lines.}
\label{fig:eigval}
\end{figure}
\begin{figure}
\centering
\includegraphics[width=0.6\textwidth]{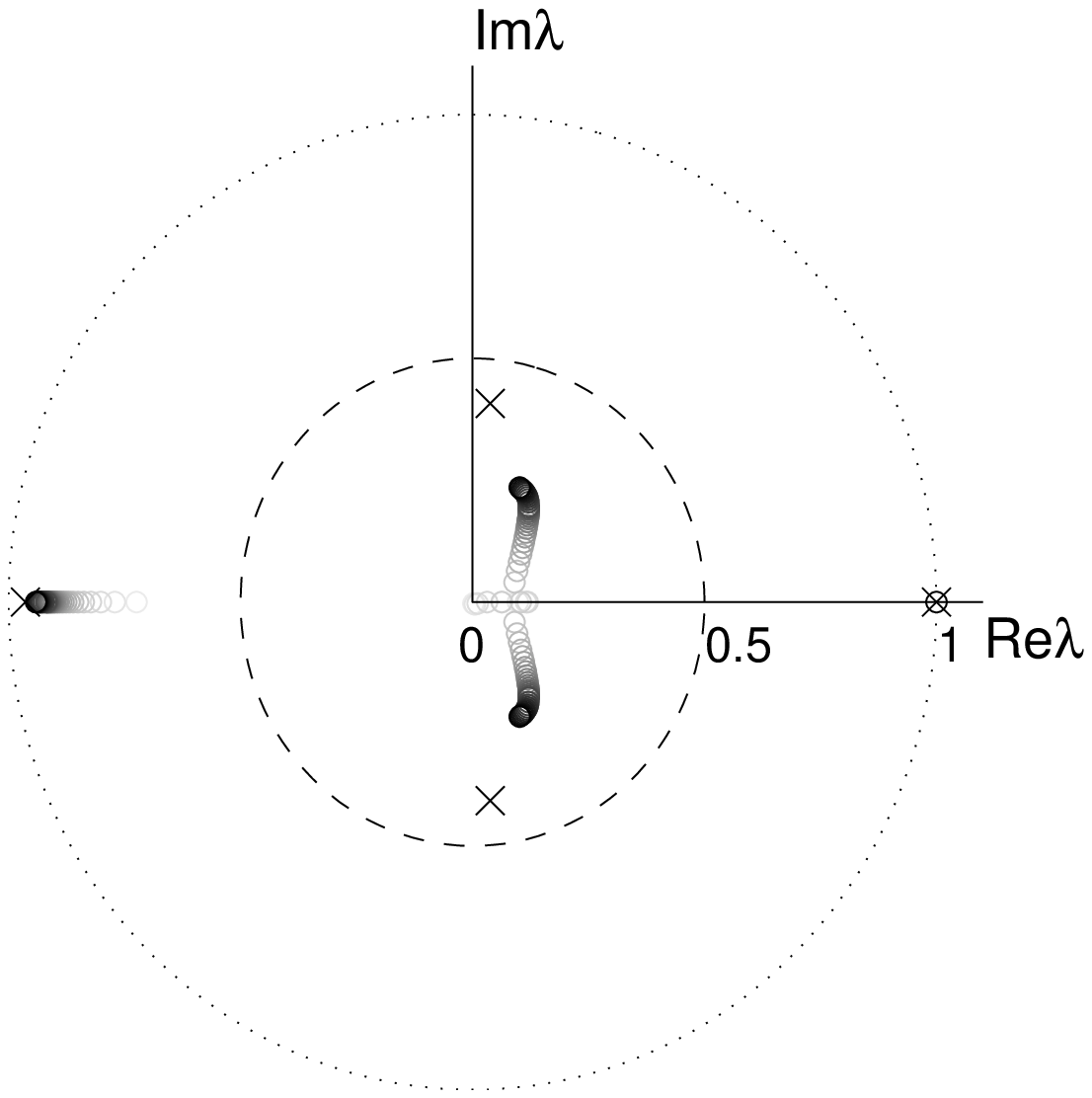}
\caption{Plot of four largest eigenvalues in the complex plane as a function
of the noise strength for $\epsilon=1/84$. The darker color, the smaller
noise, which changes from $\sigma_1\approx0.1$ to $\sigma_2\approx0.02$
($N_1=10$, $N_2=200$). Crosses represent the eigenvalues of the transition
matrix $\mathbf{T}$ with the largest moduli, while dashed circle of radius 1/2
represents the essential spectrum.}
\label{fig:4eigval}
\end{figure}
In this case
\begin{equation}
\mathbf{T} = \left(
\begin{array}{cccc}
  0  &  0  & 1/4 & 3/4\\
  0  & 1/2 & 1/8 & 3/8\\
  1  &  0  &  0  &  0 \\
 1/3 & 2/3 &  0  &  0 
\end{array}\right)\ ,
\end{equation}
and $\lambda_2=-(1+\sqrt{5})/4\approx-0.809$, while for $\epsilon=1/84$ we
have $\lambda_2\approx-0.9638$.
Fig.~\ref{fig:eigval} presents the second eigenvalue $\lambda_2$ of the
FP-operator for the system with noise obtained by numerical diagonalization of
$(N+1)$ dimensional matrix $\mathbf{D}$ as a function of the noise width
$\sigma$ for two different values of $\epsilon$. With decreasing noise
strength $\sigma$ the second eigenvalue tends to its deterministic counterpart
(represented by dashed horizontal lines). For $\epsilon=1/20$ a fair
approximation is obtained already for $N=200$ ($\sigma\approx0.02$), whereas
for $\epsilon=1/84$ we need smaller noise to obtain satisfactory results.
Fig.~\ref{fig:4eigval} shows the motion of four eigenvalues with the largest
moduli in the complex plane as the noise strength is decreased. Convergence of
the eigenvalue outside essential spectrum to its deterministic counterpart is
much faster then these of the two other eigenvalues belonging to the essential
spectrum (one may even question, whether they at all converge to the
deterministic value).
In the same way we expect that eigenvectors tend to their deterministic
counterparts.
\begin{figure}
\centering
\includegraphics[width=0.6\textwidth]{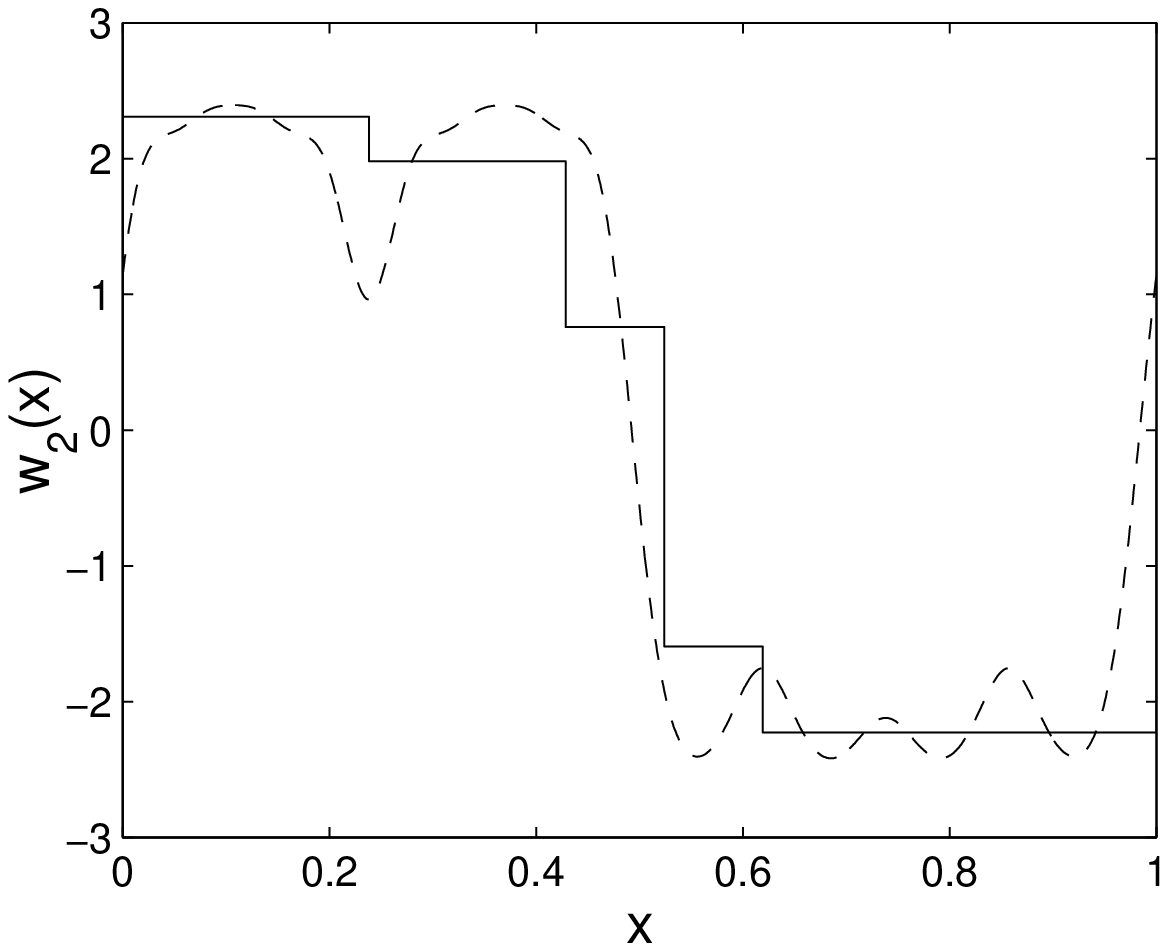}
\caption{Eigenvector corresponding to the second eigenvalue obtained from the
matrix representation of the FP-operator of noisy system ($\epsilon=1/84$) for
$\sigma\approx0.02$ (dashed line) and from the transition matrix $\mathbf{T}$
of the deterministic system (solid line).}
\label{fig:eigvec}
\end{figure}
Fig.~\ref{fig:eigvec} presents the second eigenvector obtained
from the matrix representation of the FP-operator in the noisy system
($\epsilon=1/84$) for $\sigma\approx0.02$ (dashed line) and from the
transition matrix of the Markov partition.

Eventually, we can compare the values of the subleading eigenvalue of the
spectrum with the decay rate of the autocorrelation function. The
correlation function of any function $h(x)$, defined by a time average
\begin{equation}
C_h(n) = \lim_{t\rightarrow\infty}\frac{1}{T}\sum_{t=1}^Th(x(t+n))h(x(t))
\end{equation}
is called autocorrelation function in the simplest case $h(x)=x$.
For an ergodic system it can be expressed by an average over the phase space
with respect to the invariant measure $\rho(x)$
\begin{equation}
C(n) = \int_\Omega dx\,\rho(x)\,x(n)\,x\,,\quad\mathrm{where\ }x(n) = f^n(x)\,.
\end{equation}
In the case considered, in which the dynamics is composed of two parts ---
deterministic and stochastic (see (\ref{eq:langev})) --- one has to
average additionally over different realizations of the stochastic
perturbation. Let us consider first the one--step correlation function
\begin{equation}
  C(1) = \int\langle x(f(x)+\xi)\rangle \rho(x)\,dx\ ,
\end{equation}
where the angle brackets denote averaging over different realizations of the
noise ($\xi$). Employing property (\ref{eq:expand}) we can write
\begin{eqnarray}
  C(1) &=& \int{\cal P}(f(x),x')x\,x'\,\rho(x)\,dx\,dx' =
    \sum_k\int u_k(f(x))\tilde{v}_k(x')x\,x'\,\rho(x)\,dx\,dx'\nonumber\\
    &=& \sum_k\left(\int\tilde{v}_k(x')x'\,dx'\right)
	\left(\int u_k(f(x))x\,\rho(x)\,dx\right) =
	\vec{\mathrm{v}}\cdot\vec{\mathrm{u}}\ ,
\end{eqnarray}
where we have defined $\vec{\mathrm{v}}$ and $\vec{\mathrm{u}}$ as vectors
of the integrals, $\int dx'\tilde{v}_k(x')h(x')$ and $\int
dx\rho(x)u_k(f(x))h(x)$ for $k=0,\ldots,N$. In an analogous way we obtain the
correlation function of any observable $h$ for longer delay times
\begin{eqnarray}
C_h(n) &=& \int{\cal P}(f(x_1),x_2)\ldots{\cal
    P}(f(x_n),x_{n+1})h(x_1)h(x_{n+1})\rho(x_1)\,d^{n+1}x \nonumber \\
    &=& \vec{\mathrm{v}}\,\mathbf{D}^{n-1}\vec{\mathrm{u}}\,,\quad n>0,
\end{eqnarray}
where $(N+1)$--dimensional matrix $\mathbf{D}$ is already defined in
(\ref{eq:evolution}). Thus one may expect the
correlation function to be composed of a $N+1$ exponentially decaying modes
determined by the moduli of the eigenvalues of the matrix representation of
the FP-operator.
\begin{figure}
\centering
\includegraphics[width=0.6\textwidth]{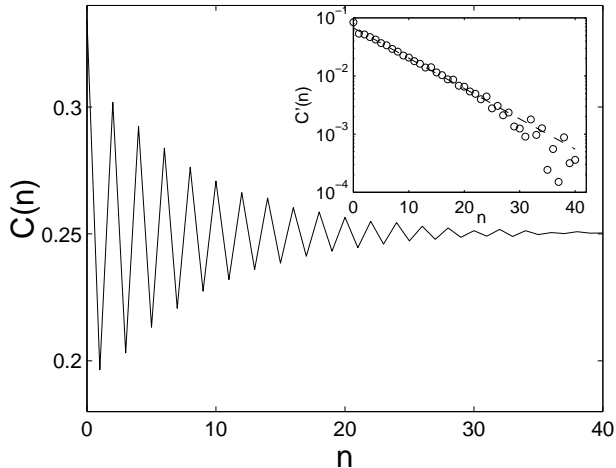}
\caption{Correlation function for map (\ref{eq:antikel_map}) with
$\epsilon=1/84$ and width of noise $\sigma\approx0.045$. The inset presents
$C'(n) = |C(n)-C(\infty)|$ in semilog scale. Exponential fit gives
$\log{C'}\approx-0.1198n$ while the subleading eigenvalue of the matrix
$\mathbf{D}$ gives $\log|\lambda_2| = -0.1201$.}
\label{fig:corr1_84}
\end{figure}
An example of such a situation is presented in Fig.~\ref{fig:corr1_84}. The
autocorrelation function is calculated by choosing $10^6$ points according to
the uniform distribution in $\Omega$ and evolving them by system $f$ given by
(\ref{eq:antikel_map}) with $\epsilon=1/84$ (and the noise characterized by
$N=50$). The value of the decay exponents obtained from the best exponential
fit (see inset in Fig.~\ref{fig:corr1_84}) agrees up to 2\% with the one we
get from the second eigenvalue of the matrix representation of the
FP-operator.

Note that the observed decay of correlation was studied for an ensemble of
points distributed uniformly in $\Omega$, that is according to the invariant
measure of the system. This seems to be more natural approach than in the case
recently analyzed by Weber \textit{et al.} \cite{WHS,WHBMS}, in which the
initial density was determined by the selected eigenvector, while the
observed decay rate was governed by the corresponding eigenvalue.

\section{Conclusions}
The aim of this paper is twofold. We point out that by introducing a
stochastic noise into a deterministic system and later tending with its
strength to zero on may distinguish a ``physically'' important part of the
spectrum of the associated FP-operator. In this way we suggest to define a
``physical'' spectrum of the classical map, as this robust with respect to
stochastic perturbations. Moreover, we provide a method for approximation of
the spectrum and eigenvectors of the FP-operator by applying a suitable noise
decomposable in the sense of (\ref{eq:expand}). This technique seems to be
more justified from the physical point of view than just
truncating the infinite dimensional matrix representation to some finite
dimension e.g.~\cite{WHS} (thus effectively introducing some perturbation),
since one constructs a system with noise of known properties and can decrease
the amplitude of the perturbation in a controlled way. Another methods of
introducing noise into deterministic systems to analyze their spectral
properties were recently used in \cite{Fishman,Agam}.

We demonstrated that the eigenvalues of the FP-operator located outside the
essential spectrum are robust not only against local perturbations as proved
in \cite{Keller}, but also against non--local perturbations of the form
(\ref{eq:trignoise}). Moreover, these stable eigenvalues have a direct
physical meaning: they determine the rate of the exponential decay of
correlation in the system. Thus our approach of analyzing dynamical system
with a stochastic perturbation of a variable strength allows one to identify
the physically important part of the FP spectrum without mathematical
ambiguities of selecting a space, in which this operator acts.

On the other hand it would be interesting to analyze spectral properties of
dynamical systems in presence of a stochastic perturbation fulfilling
properties (\ref{eq:expand}) but different than (\ref{eq:trignoise}) studied
in this paper. We expect the eigenvalues of the FP-operator not belonging to
its essential spectrum to be weakly dependent on the specific form of the
probability distribution ${\cal P}(\xi)$, but this conjecture requires
further verification.

We would like to thank W.~S\l omczy\'nski and P.~Pako\'nski for fruitful
discussions. A.O. gratefully acknowledges financial support from
Subsydium~FNP~1/99. This work was supported by a Polish KBN grant no
2~P03B~072~19.

\end{document}